\documentstyle[aps,multicol,epsf]{revtex}
\draft
\begin{document}
\title{Atom lithography with near-resonant light masks: Quantum optimization analysis}
\author{R. Arun, Offir Cohen, and I.Sh. Averbukh}
\address{Department of Chemical Physics, The Weizmann Institute of Science,
Rehovot 76100, Israel}
\date{\today}
\maketitle
\begin{abstract}
We study the optimal focusing of two-level atoms with a near
resonant standing wave light, using both classical and quantum
treatments of the problem. Operation of the focusing setup is
considered as a nonlinear spatial squeezing of atoms in the
thin- and thick-lens regimes. It is found that the near-resonant
standing wave focuses the atoms with a reduced background in
comparison with far-detuned light fields. For some parameters,
the quantum atomic distribution shows even better localization
than the classical one. Spontaneous emission effects are 
included via the technique of quantum Monte Carlo wave function 
simulations. We investigate the extent to which non-adiabatic 
and spontaneous emission effects limit the achievable minimal 
size of the deposited structures.
\end{abstract}

\begin{multicols}{2}
\section{introduction}
\vskip -0.15in
The ability to control the motion of atoms using laser fields has
led to the realization of optical elements such as mirrors, lenses,
beam splitters, etc, for atomic beams. One of the interesting
applications is the laser focusing of atoms, which is useful to the
technologically important problem of atom lithography. The principle
of atom lithography is based on using a standing wave (SW) of light
as a mask on atoms to concentrate the atomic flux periodically and
create desired patterns at the nanometer scale (for recent reviews see,
\cite{{review1},{review2}}). Since the first experimental demonstration 
of submicron atomic structures \cite{timp}, the subject has seen a 
considerable growth both
theoretically \cite{{prentis},{mcc1},{berman},{olsen},{lee}} and
experimentally \cite{{sodium},{mcc2},{chr},{alu},{ytt},{iro},{twod}}. 
In the direct deposition setup, periodic atomic lines of sodium
\cite{{timp},{sodium}}, chromium \cite{{mcc2},{chr}}, aluminum
\cite{alu}, ytterbium \cite{ytt}, and iron \cite{iro} atoms have
been successfully fabricated. The technique has also been applied to
two-dimensional pattern formation \cite{twod}.

Most theoretical studies of atom lithography employ a particle
optics approach to laser focusing of atoms
\cite{{prentis},{mcc1},{arun}}. The focal properties of the light
have been examined in terms of time dependent classical trajectories
of atoms in the light induced potential. It has been shown that the
atomic image at the focal plane exhibits a broadening due to severe
aberrations caused by anharmonicity of the sinusoidal dipole
potential \cite{{prentis},{mcc1}}. As a result, all current
lithography schemes suffer from a considerable background in the
deposited structures. A novel scheme to reduce the aberration
problem was suggested recently in \cite{arun} by using optimized
multilayer light masks. Quantum mechanical analysis of the focusing
of atomic beams has been performed as well. Cohen {\it et al.}
studied quantum mechanically the thin-lens regime of atom focusing
with both far detuned and exactly resonant standing light waves
\cite{berman}. Recently, atomic nanostructures have been
experimentally realized with exactly resonant SW \cite{resonant}. In
the thick-lens regime, focusing atoms is generally achieved with a
blue-detuned SW light whose detuning $(\Delta)$ is of the order of
the Rabi frequency $(\Omega_0)$ of the atom-light interaction
\cite{{prentis},{mcc1}}. In this case, the influence of spontaneous
emission on the focusing of atoms has been shown to be negligible by
the quantum treatment \cite{lee}. However, a detailed study of the
atom focusing with SW's that optimizes the lens performance as a
function of the detuning of the light frequency has not appeared
previously, to the best of our knowledge.

In this paper, we study atom lithography with near-resonant light
masks $(|\Delta|/\Omega_0 \lesssim 1)$ considering both blue- and
red-detuned light. We give a comprehensive theoretical analysis for
the classical and quantum treatments of the problem. The parameters
for the best focusing of atoms are found as functions of the detuning 
$(|\Delta|)$ using the optimal squeezing method developed in 
Ref. \cite{arun}. To include momentum diffusion and spontaneous emission
effects on atoms, we perform Monte Carlo wave function simulations. 
High resolution deposition of chromium atoms is considered as an
example, though the general conclusions drawn should apply well to other
atoms. 

The paper is arranged as follows. In Sec. II, the basic framework of the problem is defined
and the focusing of atoms is studied classically under the influence of adiabatic light potentials.
In this section, we examine the optimal squeezing scheme of \cite{arun} when applied to the
atomic-beam traversing a near-resonant SW light. In Sec. III, the problem is treated in the
quantum domain to account for non-adiabatic and diffraction effects on atom focusing.
The effects of spontaneous emission of atoms are considered in Sec. IV. Finally, in Sec. V,
we summarize our main results.

\section{focusing of atoms by adiabatic light potentials:  classical treatment}
We begin our discussion with a model based on the interaction of a beam of two-level atoms
with a near-resonant SW light. We take the direction of propagation of the atomic beam through
the light along the $z$ direction. The SW (assumed to be formed along the $x$ direction) has a
frequency of $\omega_l$ and $\Delta = \omega_l - \omega_0$ defines the detuning of the light
frequency from the atomic transition frequency $\omega_0$. The atom-light interaction is
characterized by the Rabi frequency
\begin{equation}
\Omega(x,z) = \Omega_0 \exp(-z^2/\sigma_z^2) \cos(kx)~.
\end{equation}
Here, the term $\exp(-z^2/\sigma_z^2)$ accounts for the spatial
variation (a Gaussian beam profile with diameter $\sigma_z$) of the
light intensity along the $z$ direction. The $cos(kx)$ term comes
from the sinusoidal $[I(x) \propto \cos^2(kx)]$ variation of the SW
intensity along the $x$ direction. The quantity $\Omega_0$
represents the peak Rabi frequency of the atom-light interaction and
$\lambda = 2\pi/k$ is the wavelength of the laser beams forming the SW.
The velocity $v_z$ of the atoms along the beam axis is sufficiently
large, so that the atom's position along the $z$ direction can be
replaced by the time dependence $z = v_z t$. Defining $\sigma_t =
\sigma_z/v_z$, the time dependent Rabi frequency is thus given by
\begin{equation}
\Omega(x,t) = \Omega_0 \exp(-t^2/\sigma_t^2) \cos(kx)~.
\label{rabi}
\end{equation}

The behavior of atoms in the near-resonant light field can be best
understood in the dressed state picture of the atom-light
interaction \cite{dress}. The dressed states, which are the
eigenstates of the interaction Hamiltonian, depend on time through
the time-dependent Rabi coupling Eq. $(\ref{rabi})$. If the
Hamiltonian temporal variation is smooth, the atom prepared
initially in one of the eigenstates of the Hamiltonian will follow
the time-dependent eigenstate. The corresponding adiabatic condition
is \cite{adiab}
\begin{equation}
|\Delta| \gg \sqrt{\frac{\Omega_0}{\sigma_t}}~. \label{adia}
\end{equation}
We assume that the atoms in the beam are initially in their ground
state and that the adiabatic condition Eq. $(\ref{adia})$ is
satisfied. In this case, the atoms can be described as pointlike
particles moving in  the  potential
\begin{equation}
U(x,t) = Sgn(\Delta) \frac{\hbar}{2} \sqrt{\Delta^2 + \Omega^2(x,t)}~.
\label{pot}
\end{equation}
Here $Sgn(\Delta) = +1~(-1)$ for $\Delta > 0$ $(\Delta < 0)$. Note
that this adiabatic potential should be contrasted with the light
shift felt by the bare atomic states in the far-detuned limit
\cite{berman}. The adiabatic potentials and the focusing of atoms
are shown schematically in Fig. 1.

Many aspects of atom focusing by the standing light waves can be explained in the semiclassical
picture of the atom's interaction with the light induced potential
\cite{{prentis},{mcc1},{arun}}. Therefore, we start
considering the problem in the classical framework. We neglect spontaneous emissions from the
atoms by assuming that the atom's interaction time with the light is much shorter compared
to the lifetime of the excited atomic level. The classical trajectories of atoms in the
adiabatic light potential Eq. $(\ref{pot})$ obey Newton's equation of motion
\begin{equation}
\frac{d^2x}{dt^2} + \frac{1}{m} \frac{\partial U(x,t)}{\partial x} = 0 ~,
\label{newton}
\end{equation}
\vskip -0.2 in
\begin{figure}[t]
\epsfxsize=210pt
\centerline{
\epsfbox{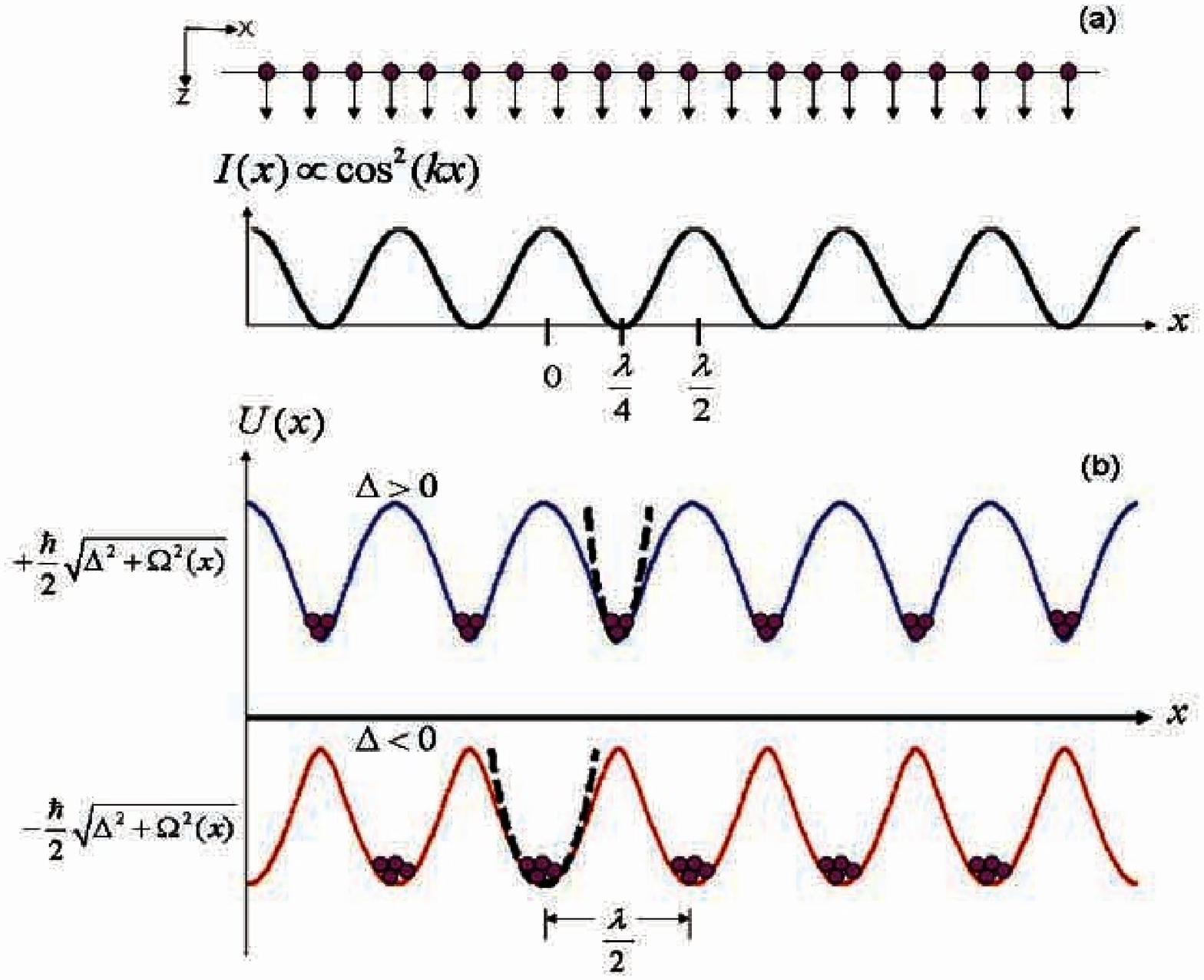}
}
\caption{(Color online) Schematic representation of laser focusing of atoms by a SW light.
(a) A collimated atomic beam impinges on the near-resonant SW light whose intensity varies
sinusoidally along the $x$ direction. (b) The interaction with the light induces the adiabatic
potential [either plus or minus in Eq. (\ref{pot})] for the external motion of atoms depending
upon the sign of the detuning. For a positive (blue) detuning, the atoms localize near the minima
of the light intensity, while for a negative (red) detuning they remain localized near the
light intensity maxima. The dashed curves represent the quadratic approximation to the
adiabatic potentials near their minima.}
\end{figure}
\vskip 0.13in
\noindent 
where $m$ is the atomic mass. The force induced by the SW focuses (localizes) the atoms near
the intensity minima (maxima) for $\Delta > 0$ $(\Delta < 0)$. As a measure of the atomic
localization, we use the localization factor \cite{arun}
\begin{eqnarray}
L(t) &=& 1 + Sgn(\Delta) \langle \cos[2kx(t,x_0)] \rangle \nonumber \\ 
&\equiv& \frac{2}{\lambda} \int_{-\lambda/4}^{\lambda/4} dx_0 
\{1 + Sgn(\Delta) \cos[2kx(t,x_0)]\}~, \label{local}
\end{eqnarray}
where $x(t,x_0)$ is the solution of the differential equation
$(\ref{newton})$ satisfying the initial condition $x \rightarrow
x_0$ at $t \rightarrow -\infty$. The average in Eq. $(\ref{local})$
is taken over the random initial positions of the atoms, and the
localization factor is measured as a function of time $t$ counted
from the moment $(t=0)$ of passing the Gaussian center  of the SW.
The localization factor equals zero for an ideally localized atomic
ensemble and is proportional to the mean-square variation of the $x$
coordinate (modulo standing-wave period) in the case of a
well-localized distribution $(L\ll1)$.

The localization factor defined in Eq. $(\ref{local})$ measures a
nonlinear spatial focusing of atoms beyond the linear paraxial
approximation. The plane of the best atomic localization (minimal
atomic background) is determined by minimizing the localization
factor \cite{arun}. This optimization can be done either in the
thin- or the thick-lens limit of atom focusing. First we consider
the thin-lens focusing \cite{pfau} of atoms by the adiabatic potential
Eq. $(\ref{pot})$, which 
\begin{figure}[t]
\centerline{ \epsfxsize=185 pt \epsfbox{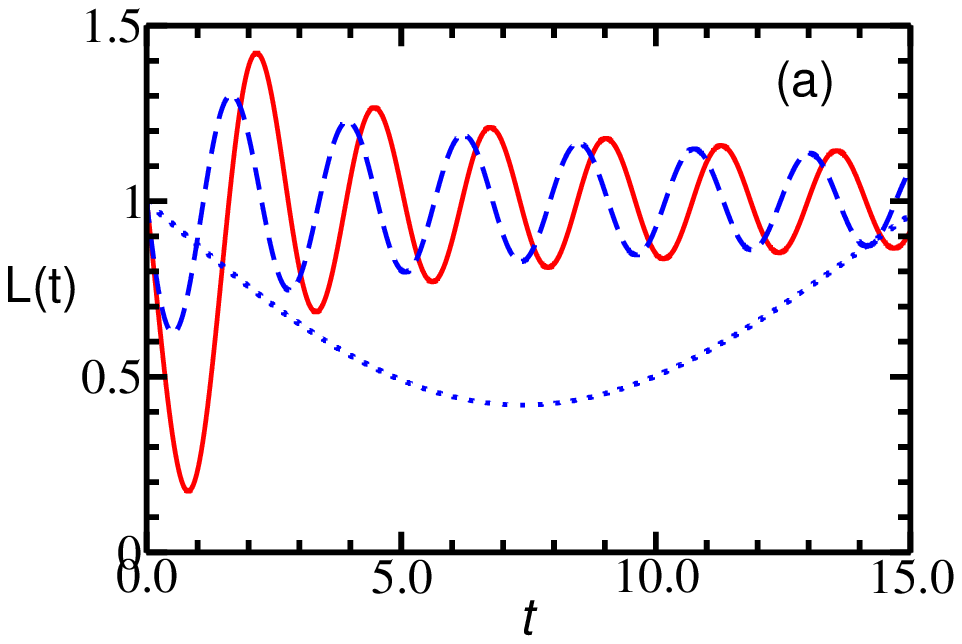}}
\vskip 0.08in \centerline{\epsfxsize=185 pt \epsfbox{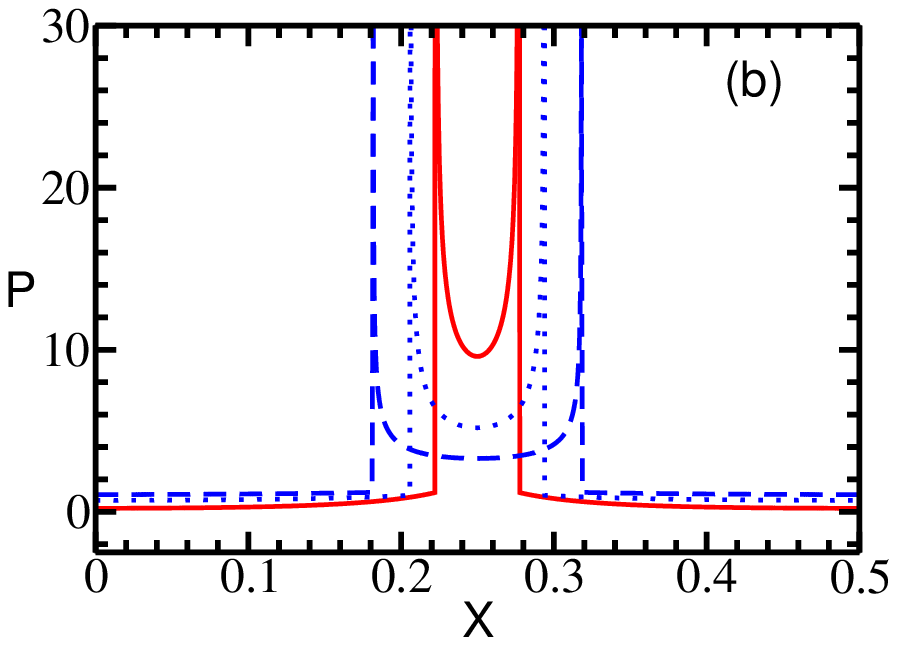}}
\vskip 0.03in
\caption{(Color online) (a) Localization factor of the atomic distribution as
a function of the dimensionless time $t$ for $\sigma_t = 0.07$ and
$\Delta/\Omega_0 = -0.125$ (solid curve), $\Delta/\Omega_0 = 0.125$ (dashed curve),
$\Delta/\Omega_0 = 5$ (dotted curve). The global minimum values of $L(t)$
are 0.17 (solid curve), 0.63 (dashed curve), and 0.42 (dotted curve).
(b) Probability density (P) of the atomic distribution at the time $t = t_m$
of the best atomic localization. The parameters are same as those of (a) with
$t_m = 0.82$ (solid curve), $t_m = 0.52$ (dashed curve), and $t_m = 7.38$
(dotted curve). For the sake of comparison, the solid curve has been displaced
by 0.25 units along the $X$ axis. The times $t_m$ correspond to the global
minima of the localization factor in (a).}
\end{figure}

\vskip 0.15in
\noindent
is valid in the Raman-Nath approximation to the atom-light interaction.
In this case, the atomic displacement along
the SW direction is negligible within the light and the focal point
is well outside the region of the light fields \cite{mcc1}.
Within thin-lens approximation, the SW introduces a spatially
dependent sudden kick for atoms along the transverse ($x$ axis)
direction. For atoms with zero initial velocity along the $x$ axis,
the change in velocity can be calculated from Eq. $(\ref{newton})$
to be
\begin{eqnarray}
\delta v_x &=& Sgn(\Delta) \frac{\hbar k \Omega_0^2}{4 m} \nonumber \\ 
&&~\times \int dt \frac{\exp(-2t^2/\sigma_t^2)\sin[2kx(t)]}{\sqrt{\Delta^2 + 
\Omega_0^2 \exp(-2t^2/\sigma_t^2) \cos^2[kx(t)]}}~, \label{xapprox}
\end{eqnarray}
where the integration is done over the duration $(\pm 3\sigma_t)$ of
the Gaussian profile. In the Raman-Nath approximation, the position
$x(t)$ of the atom in Eq. ($\ref{xapprox}$) during interaction with
the light can be approximated to be its initial position $x_0$.
After passing through the light region, the
\vskip -0.18in 
\begin{figure}[t]
\centerline{ \epsfxsize=180 pt \epsfbox{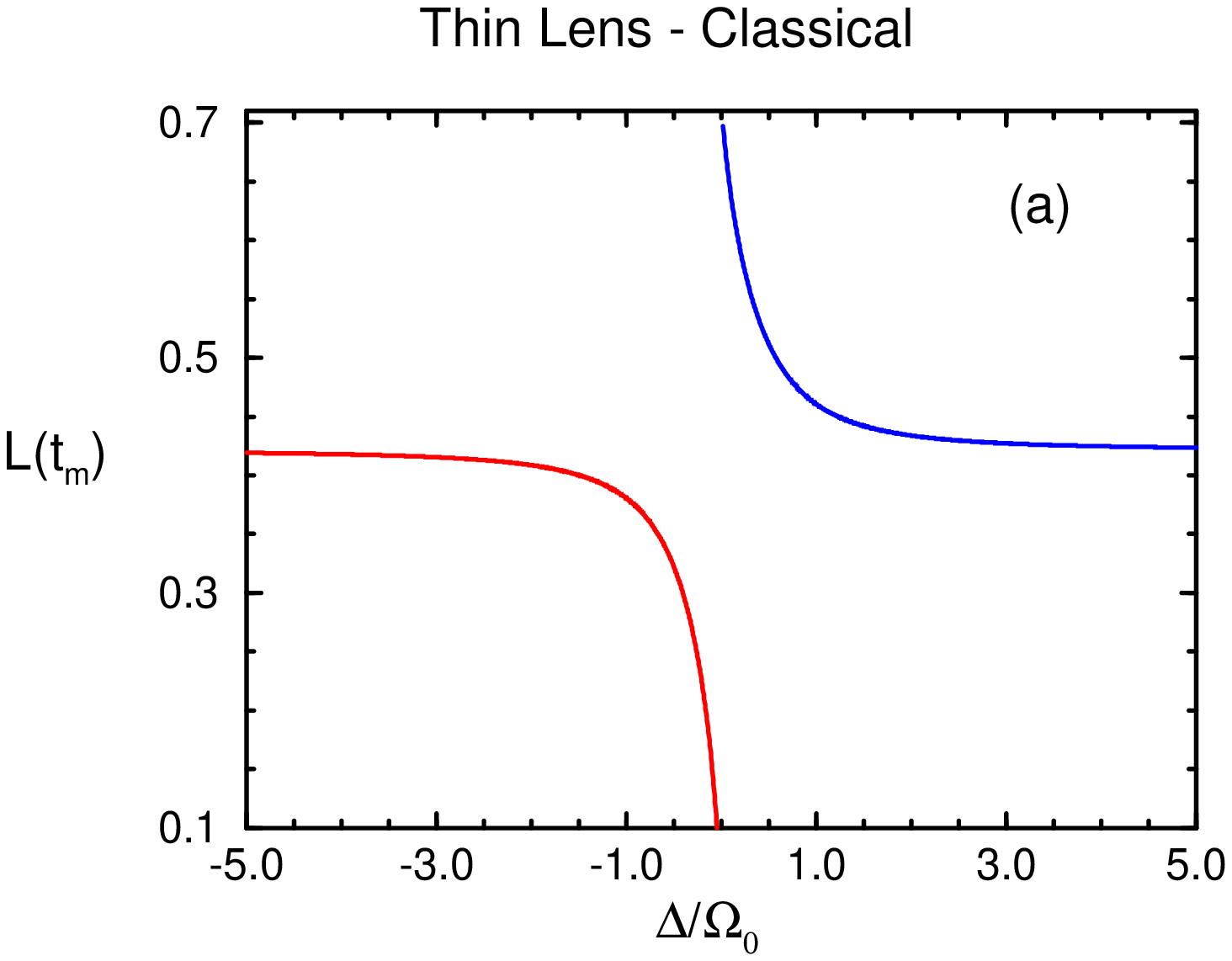}}
\vskip 0.08in \centerline{\epsfxsize=180 pt \epsfbox{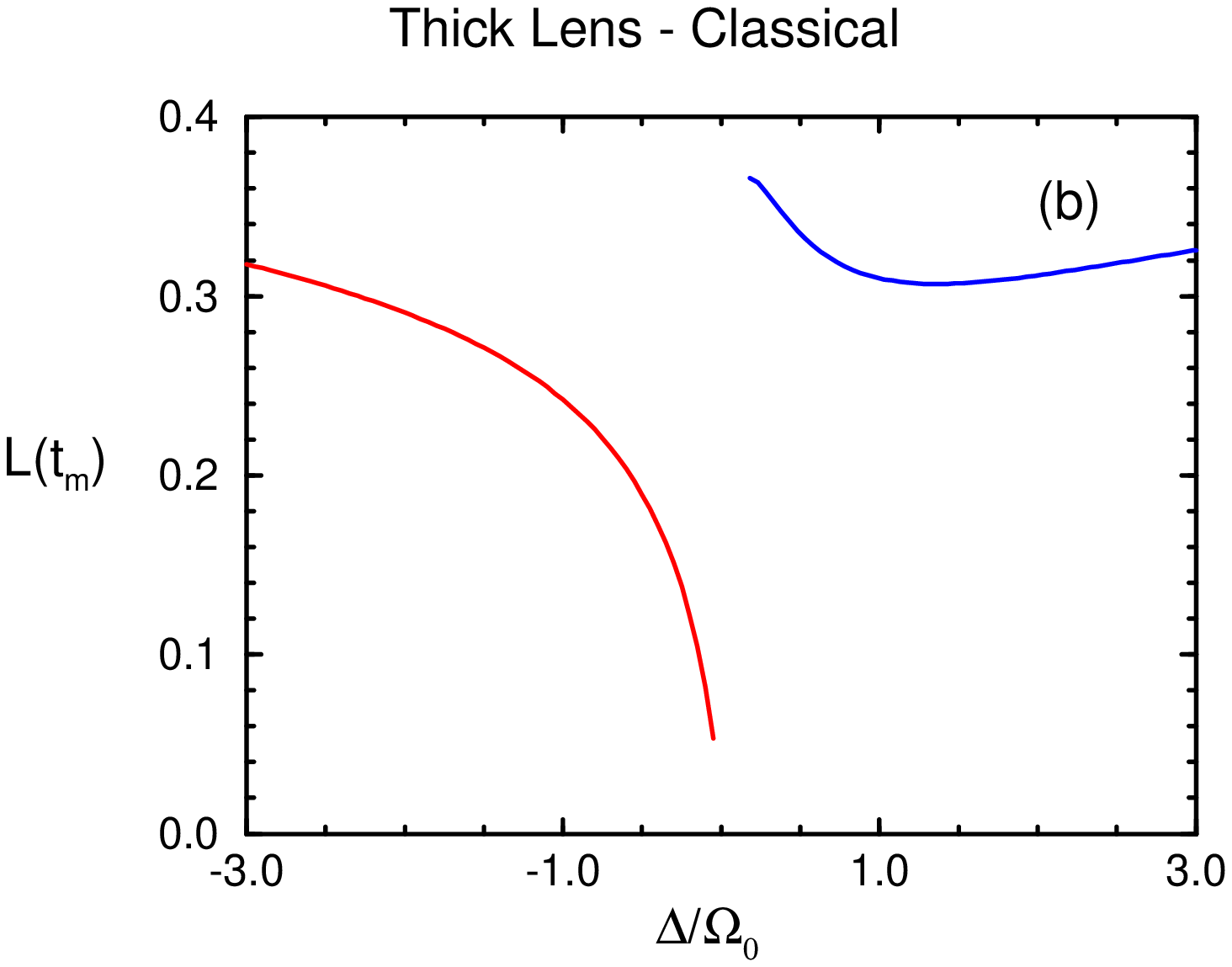}}
\vskip 0.03in
\caption{(Color online) Minimal localization factor of the atomic distribution 
as a function of the detuning $\Delta/\Omega_0$ for (a) $\sigma_t = 0.07$, 
(b) $\sigma_t = 4$.}
\end{figure}

\vskip 0.05in
\noindent
atom moves as a free particle with a time dependent transverse position
\begin{equation}
x(t,x_0) = x_0 + \delta v_x t~. \label{pos}
\end{equation}
Using Eqs. (\ref{local}) and (\ref{pos}), we calculate the
localization factor  as a function of dimensionless time for both
blue $(\Delta > 0)$ and red $(\Delta < 0)$ detuning conditions. The
results are shown in Fig. 2 with plots of the atomic distribution
\cite{norm} at the time of minimal localization factor. These
results are also compared with the best atomic localization that can
be achieved with the usual far-detuned $(\Delta/\Omega_0 \gg 1)$ SW
light. We use dimensionless variables in which position is measured
in units of the optical wavelength $(\lambda)$, frequency in units
of the recoil frequency $(\omega_{rec} \equiv \hbar k^2/2 m)$, and
time in units of $1/[\omega_{rec}\Omega_0\sigma_t]$. It is seen from
the graph that the near-resonant light with red detuning from the
atomic transition focuses the atoms better  compared with the
far-detuned light. The atomic localization improves  by decreasing
the magnitude of the detuning in the case of atom focusing by the
red-detuned light. This is shown in Fig. 3(a), where we plot the
best localization factor as a function of the detuning parameter
$(\Delta/\Omega_0)$. Note that the localization factor saturates to
the well known value of $L \approx 0.42$ in the far-detuned
$(|\Delta|/\Omega_0 \gg 1)$ case \cite{arun}. It tends to the
asymptotic value of $L\approx0.1$ for $\Delta \rightarrow 0^{-}$. We
note, however, that nonadiabatic quantum effects should be taken
into account when the adiabatic condition, Eq. (\ref{adia}) fails
(see the next section).

\vskip -0.1in
\begin{figure}[t]
\centerline{ \epsfxsize=185 pt \epsfbox{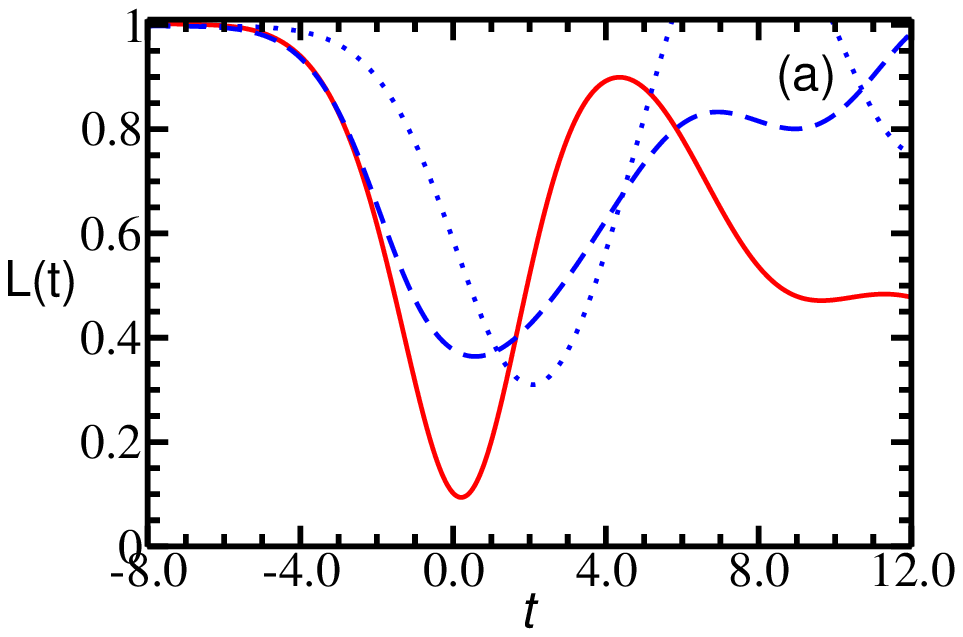}}
\vskip 0.08in \centerline{\epsfxsize=185 pt \epsfbox{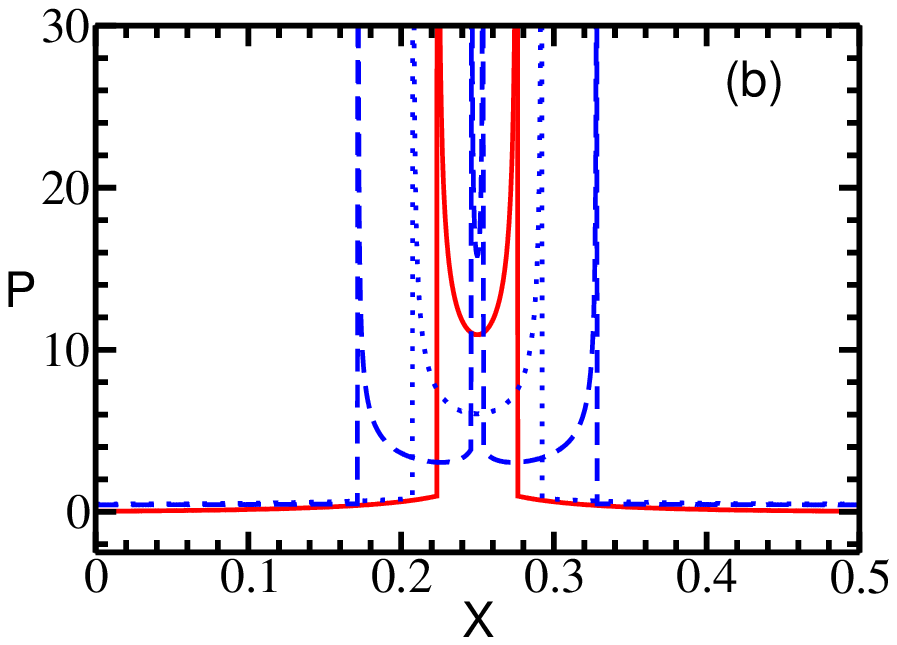}}
\vskip 0.03in
\caption{(Color online) (a) Localization factor of the atomic distribution as
a function of the dimensionless time $t$ for $\sigma_t = 4$, and $\Delta/\Omega_0 =
-0.125$ (solid curve), $\Delta/\Omega_0 = 0.125$ (dashed curve), $\Delta/\Omega_0 =
1$ (dotted curve). The global minimum values of $L(t)$ are 0.1 (solid curve),
0.36 (dashed curve), and 0.31 (dotted curve). (b) Probability density (P) of the
atomic distribution at the time $t = t_m$ of the best atomic localization. The
parameters are same as those of (a) with $t_m = 0.2$ (solid curve), $t_m = 0.58$
(dashed curve), and $t_m = 2$ (dotted curve). For the sake of comparison, the solid
curve has been displaced by 0.25 units along the $X$ axis. The times $t_m$
correspond to the global minima of the localization factor in (a).}
\end{figure}
\vskip 0.21in
In the opposite limit of the thin lens, the atoms are focused within
the region of the light fields \cite{mcc1}. The atom focusing in
this limit (known as the thick-lens focusing) is very similar to the
operation of the graded index lens in traditional optics. Neglecting
the time dependence in the Rabi frequency, we find the focal length
in the paraxial approximation as
\begin{eqnarray}
f &=&
\frac{v_z\pi}{2\Omega_0}\sqrt{\frac{\Delta}{\omega_{rec}}}~~~~~~~~~~~~~[\Delta
> 0,
\hbox{Thick lens}]~, \nonumber \\
f &=& \frac{v_z\pi}{2\Omega_0} \sqrt{\frac{\sqrt{\Delta^2 +
\Omega_0^2}}{\omega_{rec}}} ~~~~~ [\Delta < 0, \hbox{Thick lens}]~.
\end{eqnarray}
From the above expression, it is seen that the focal length is
greater for red detuning. To go beyond the paraxial approximation,
the solution $x(t)$ of the atomic motion needs to be obtained
directly by the numerical integration of Eq. (\ref{newton}). Using Eqs.
(\ref{newton}) and (\ref{local}), we obtain the localization factor
of the atomic distribution as shown in Figs. 3(b) and 4.

On comparing the graphs from Figs. (2) to (4), we see a qualitative
similarity between the thin- and thick- lens focusing of atoms. In
particular, focusing atoms by the red-detuned light gives rise to a
much reduced background of deposited atoms in comparison to the case
of atom focusing by the blue-detuned light. In the paraxial 
approximation, this can be explained by expanding the adiabatic 
potentials Eq. (\ref{pot}) upto the quadratic terms (parabolic fitting) 
near the potential minima \cite{jurgens}. As shown in Fig. 1, the 
spatial range, in which the quadratic approximation is valid, is wider 
for the case of negative detuning $(\Delta)$, which results in the 
reduced aberrations. We note that the above analysis is valid only in 
the range of parameters for which the adiabatic condition 
Eq. (\ref{adia}) is satisfied. For smaller detunings 
$(|\Delta|/\Omega_0 \ll 1)$, non-adiabatic and quantum effects may 
dominate which will be examined in the next section.

\section{optimal atomic squeezing - effects of the wave nature of atoms}
In the classical treatment of atom focusing discussed so far, the
internal structure of the atom was completely ignored and the atomic
motion in single adiabatic potential was studied in the time
domain. This procedure is valid even if the atomic motion is treated
quantum mechanically, provided the adiabatic condition (\ref{adia})
is satisfied. In the quantum treatment, the atomic center-of-mass
(c.m.) wave function $\psi(x,t)$ evolves in time according to a
Schr\"{o}dinger equation in which the potential $U(x,t)$ is given by
Eq. (\ref{pot}):
\begin{equation}
i \hbar \frac{\partial}{\partial t} \psi(x,t) = \left[\frac{p_x^2}{2m} + U(x,t)\right] \psi(x,t)~,
\label{single}
\end{equation}
where  $p_x$ denotes the c.m. momentum operator of the atom along
the SW ($x$ axis) direction.

While the time evolution, Eq. (\ref{single}) is useful to study the
quantum effects on atom focusing, the situation becomes more
complicated if the light detuning  is relatively small $(|\Delta|
\ll \Omega_0)$. In this case, non-adiabatic effects arise from
transitions between the dressed atomic states. Therefore, in order
to cover uniformly a wide range of detunings, we consider the
evolution of the atomic wave function directly in the bare-states
basis, first neglecting spontaneous emission from the atoms. The
Hamiltonian for a two-level atom with excited ($|e\rangle$) and
ground ($|g\rangle$) states interacting with the SW light is given
by
\begin{eqnarray}
H(t) &=& \frac{p_x^2}{2m} - \frac{\hbar \Delta}{2} (|e\rangle\langle e| - |g\rangle \langle g|) 
\nonumber \\ 
&& + \frac{\hbar \Omega(x,t)}{2} (|e\rangle\langle g| + |g\rangle \langle e|)~,\label{ham}
\end{eqnarray}
where $\Omega(x,t)$ is given by Eq. (\ref{rabi}).

The wave function of the two-level atom may be expressed as
\vskip -0.18in
\begin{figure}[t]
\centerline{ \epsfxsize=180 pt \epsfbox{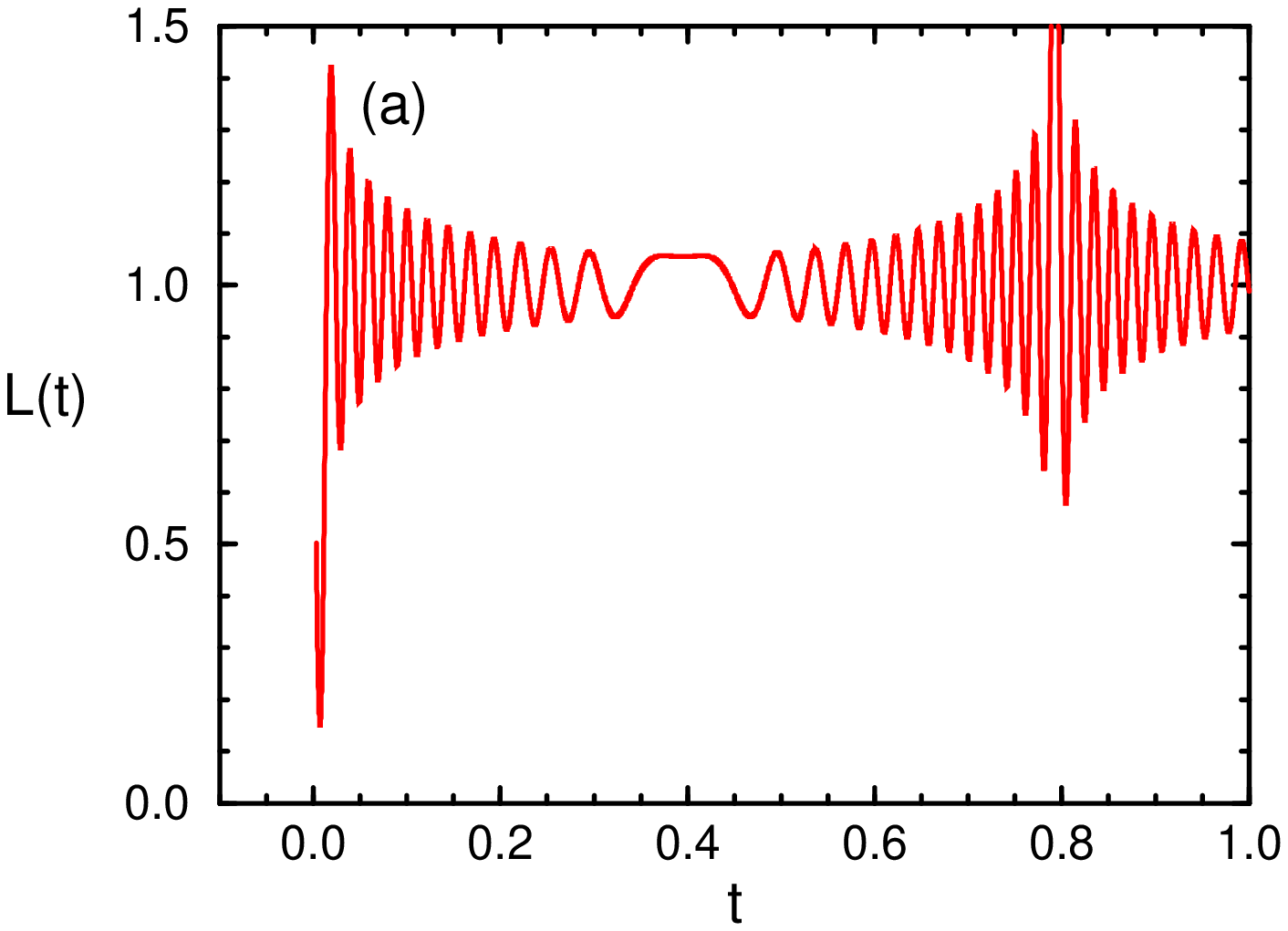}}
\vskip 0.04in \centerline{\epsfxsize=180 pt \epsfbox{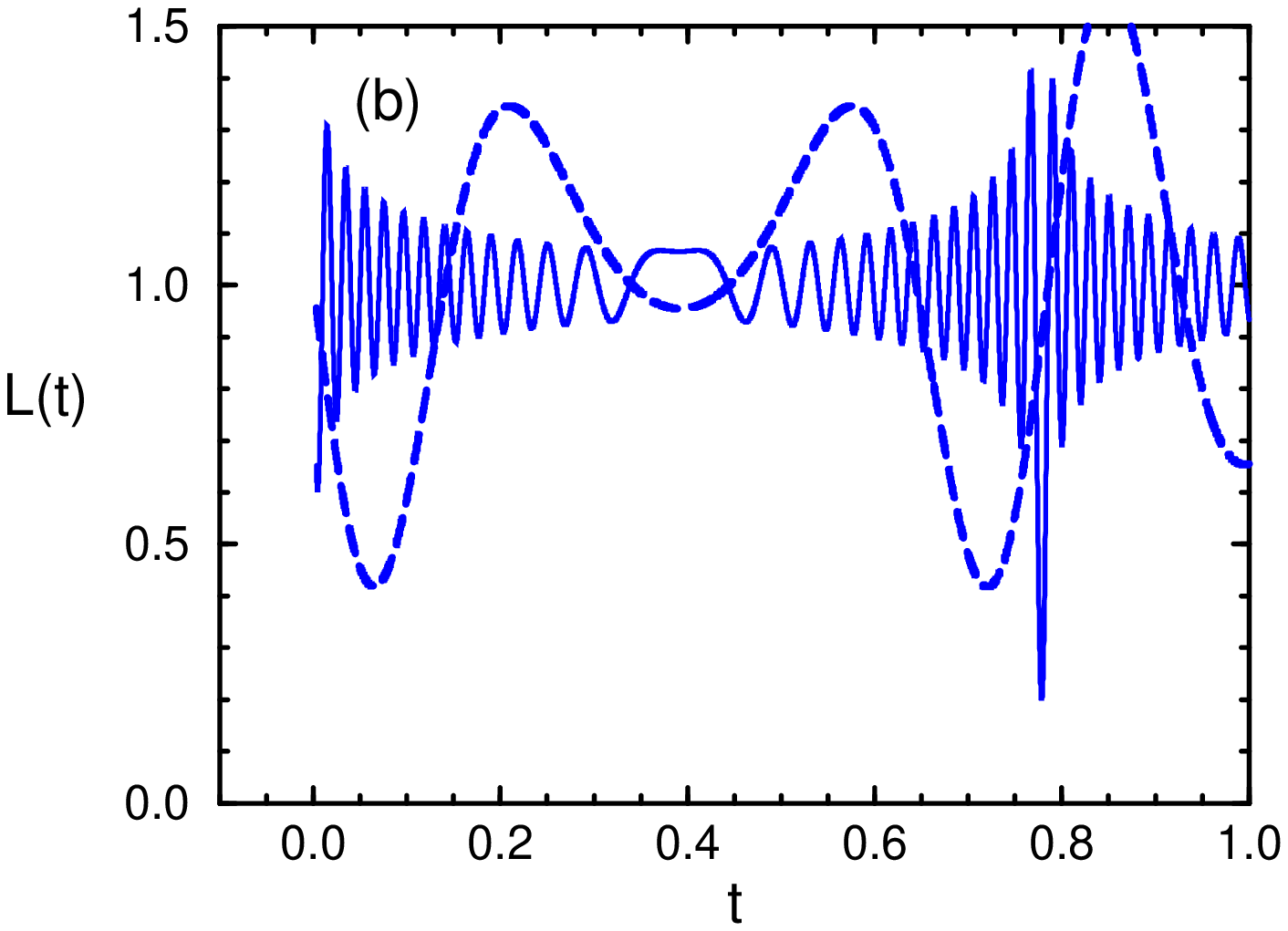}}
\vskip 0.03in
\caption{(Color online) Localization factor of the atomic distribution as
a function of the dimensionless time $t$ for the parameters $\sigma_t = 0.0006$,
$\Omega_0 = 1.92 \times 10^5$, and (a) $\Delta/\Omega_0 = -0.125$,
(b) $\Delta/\Omega_0 = 0.125$ (solid curve), $\Delta/\Omega_0 = 5$ (long-dashed curve).
The global minimum values of $L(t)$ are (a) 0.15, and (b) 0.2 (solid curve),
0.42 (long-dashed curve).}
\end{figure}

\vskip 1.5in
\begin{equation}
\Psi(x,t) = {\psi_e(x,t) \choose \psi_g(x,t)}~. \label{wave}
\end{equation}
Here $\psi_{e,g}(x,t)$ correspond to the c.m. wave functions of the
atom in its excited and ground states. We consider a spatially
uniform beam of ground state atoms having initially  zero momentum
along the SW direction. The initial wave function of the atom
normalized over the region of the SW period \cite{norm} is then given by
$\psi_g(x,t_0) = \sqrt{2/\lambda}$, where the initial time $t_0
\rightarrow -\infty$. Since the atomic distribution is expected to
be periodic (in space) after interaction with the light field, the
wave functions at time $t$ can be Fourier expanded as
\begin{eqnarray}
\psi_e(x,t) &=& \sum_{n=-\infty}^{n=\infty} C^{e}_n(t)~e^{i(2n+1)kx}~, \nonumber \\
\psi_g(x,t) &=& \sum_{n=-\infty}^{n=\infty} C^{g}_n(t)~e^{i2nkx} ~. \label{expand}
\end{eqnarray}
The Fourier coefficients $C^{e}_n(t)$ and $C^{g}_n(t)$, defined above, represent the probability
amplitudes for finding the atom in the excited and ground states with momentum $(2n+1)\hbar k$
and $2n\hbar k$, respectively. Using Eqs. (\ref{ham})-(\ref{expand}), the Schr\"{o}dinger equation
$i \hbar \partial \Psi/\partial t = H \Psi$ then leads to coupled equations for the Fourier amplitudes
as
\begin{figure}[t]
\centerline{ \epsfxsize=185 pt \epsfbox{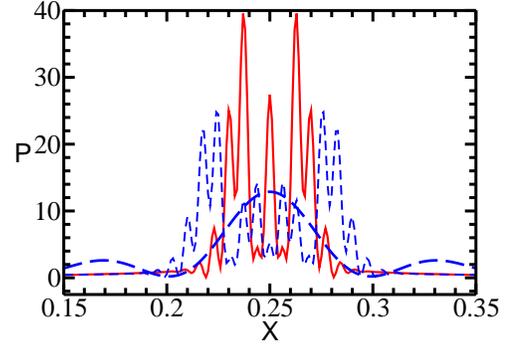}}
\vskip 0.03in
\caption{(Color online) Probability density $(P = |\Psi(x,t)|^2)$ of the atomic distribution 
at the time $t = t_m$ of the best atomic localization. The parameters are $\sigma_t = 0.0006$,
$\Omega_0 = 1.92 \times 10^5$, and $\Delta/\Omega_0 = -0.125$, $t_m = 7 \times 10^{-3}$
(solid curve), $\Delta/\Omega_0 = 0.125$, $t_m = 0.778$ (dashed curve),
$\Delta/\Omega_0 = 5$, $t_m = 0.72$ (long-dashed curve). For the sake of
comparison, the solid curve has been displaced by 0.25 units along the $X$ axis.
The times $t_m$ correspond to the global minima of the localization factor in Fig. 5.}
\end{figure}

\vspace*{0.15in}
\begin{eqnarray}
i \frac{d}{dt} C^{e}_n(t) &=& \left\{(2n+1)^2 \omega_{rec} - \frac{\Delta}{2}\right\}
C^{e}_n(t) \nonumber \\
&& + \frac{\Omega(t)}{4}(C^{g}_{n}(t) + C^{g}_{n+1}(t))~, \nonumber \\
i \frac{d}{dt} C^{g}_n(t) &=& \left\{4~n^2 \omega_{rec} + \frac{\Delta}{2} \right\}
C^{g}_n(t) \label{couple} \\ 
&& + \frac{\Omega(t)}{4}(C^{e}_{n}(t) + C^{e}_{n-1}(t))~, \nonumber 
\end{eqnarray}
with $\Omega(t) = \Omega_0 \exp(-t^2/\sigma_t^2)$.

We now proceed to the calculation of the quantum localization factor defined as
\begin{eqnarray}
L(t) &=& 1 + Sgn(\Delta) \langle \cos(2kx) \rangle \nonumber \\
&\equiv& 1 + Sgn(\Delta) \int_{-\lambda/4}^{\lambda/4} dx {|\Psi(x,t)|}^2 \cos(2kx)~.
\label{qlocal}
\end{eqnarray}
The atomic density ${|\Psi(x,t)|}^2 = {|\psi_e(x,t)|}^2 +
{|\psi_g(x,t)|}^2$ is found using Eqs. (\ref{expand}). For the
initial beam of ground state atoms $(C^{g}_0(-\infty) =
\sqrt{2/\lambda})$, the localization factor can be obtained by
solving numerically Eqs. (\ref{couple}):
\begin{eqnarray}
L(t) &=& 1 + Sgn(\Delta) \frac{\lambda}{2} \nonumber \\
&&\times\hbox{Re} \sum_{n=-\infty}^{n=\infty}
[C^e_n(t) C^e_{n+1}(t)^* + C^g_n(t) C^g_{n+1}(t)^*].\label{lval}
\end{eqnarray}
As discussed in the previous section, the best spatial squeezing (localization) of
atoms occurs at the time $t=t_m$ of the global minimum of the localization factor.
To minimize the localization factor Eq. (\ref{lval}), we first rescale the time variable
$t$ in terms of the recoil time $t_{rec} \equiv 1/\omega_{rec}$. In the classical analysis
of atom focusing, 
\vskip -0.18in
\begin{figure}[t]
\centerline{ \epsfxsize=180 pt \epsfbox{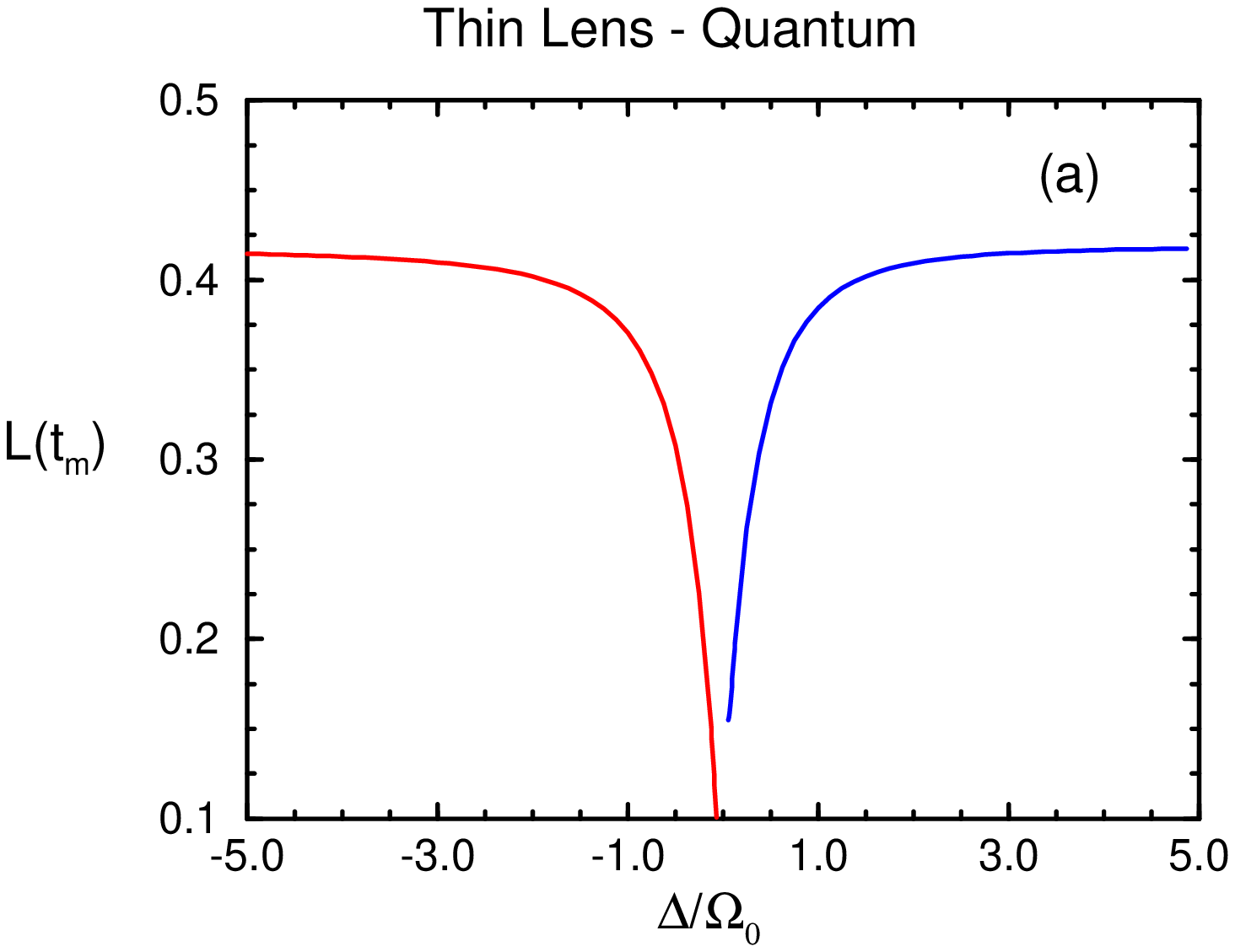}}
\vskip 0.08in \centerline{\epsfxsize=180 pt \epsfbox{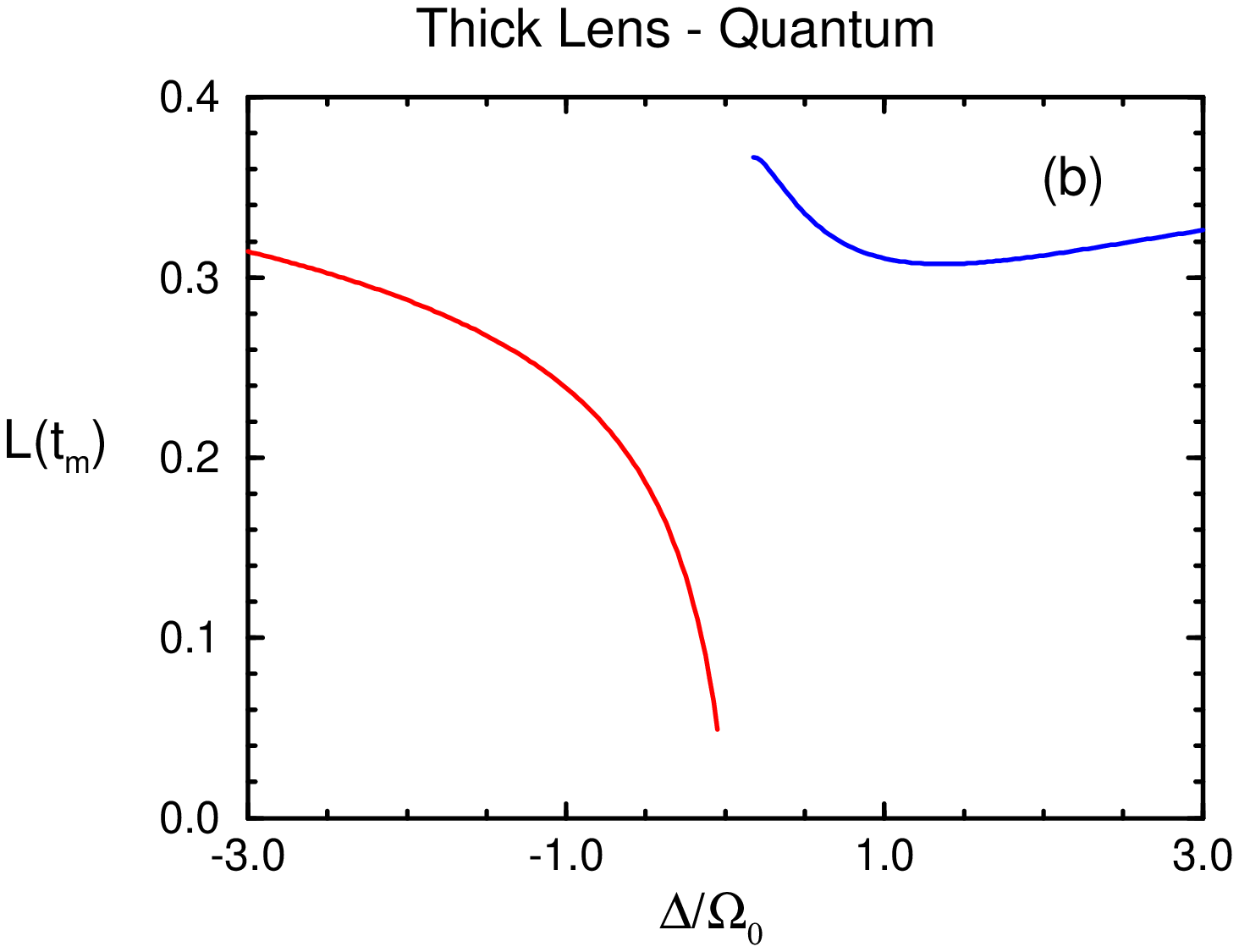}}
\vskip 0.03in
\caption{(Color online) Minimal localization factor of the atomic
distribution as a function of the detuning $\Delta/\Omega_0$ for
(a) $\sigma_t = 0.0006$, (b) $\sigma_t = 0.01$.}
\end{figure} 

\vspace*{-0.03in}
\noindent
we used a time scale that depends on the strength $(\Omega_0)$ of the
atom-light interaction. However, a quantized space-periodic motion of ground state atoms in
free space \cite{ground} repeats itself (in time) after each revival period of $t_R = \pi t_{rec}/2$.
It is therefore convenient to use the recoil time $t_{rec}$ for scaling the variables $t$ and
$\sigma_{t}$.

In Figs. 5 and 6, we display the localization factor and the atomic
spatial density for the thin-lens focusing of atoms. On comparing
the quantum mechanical localization factor with its classical
counterpart (see Figs. 2 and 5), we see that they have the same
structure for short times and the same minimal values for both blue
and red detuning situations. However, the localization factor
differs considerably in the long time evolution of the atomic
distribution. It reaches a new minimum value of $L = 0.2$ at about
half of the revival period for the blue detuned $(\Delta/\Omega_0 =
0.125)$ light. Due to these distinctly quantum features in the atom
localization, the quantum atomic distribution becomes narrower than
the classical distribution (compare Figs. 2(b) and 6) in the atom
focusing by the blue detuned light. Next, we show the best quantum
localization of atoms that can be achieved by varying the detuning
in Fig. 7(a). Again comparing it with the classical result (see Fig.
3(a)), it is seen that they are  identical for the red detuning
$(\Delta < 0)$ case. New quantum features exist only in the focusing
of atoms by the blue detuned light $(\Delta > 0)$. For the
thick-lens focusing of atoms, the results are shown in Figs. 7(b)
and (8). In this case, the quantum and classical results are quite 

\begin{figure}[t]
\centerline{ \epsfxsize=185 pt \epsfbox{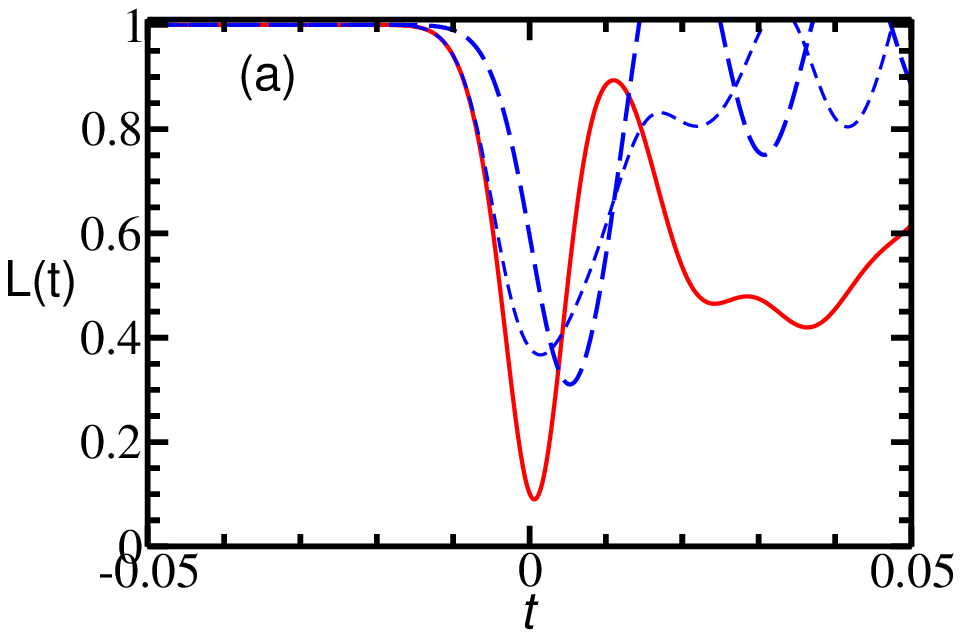}}
\vskip 0.08in \centerline{\epsfxsize=185 pt \epsfbox{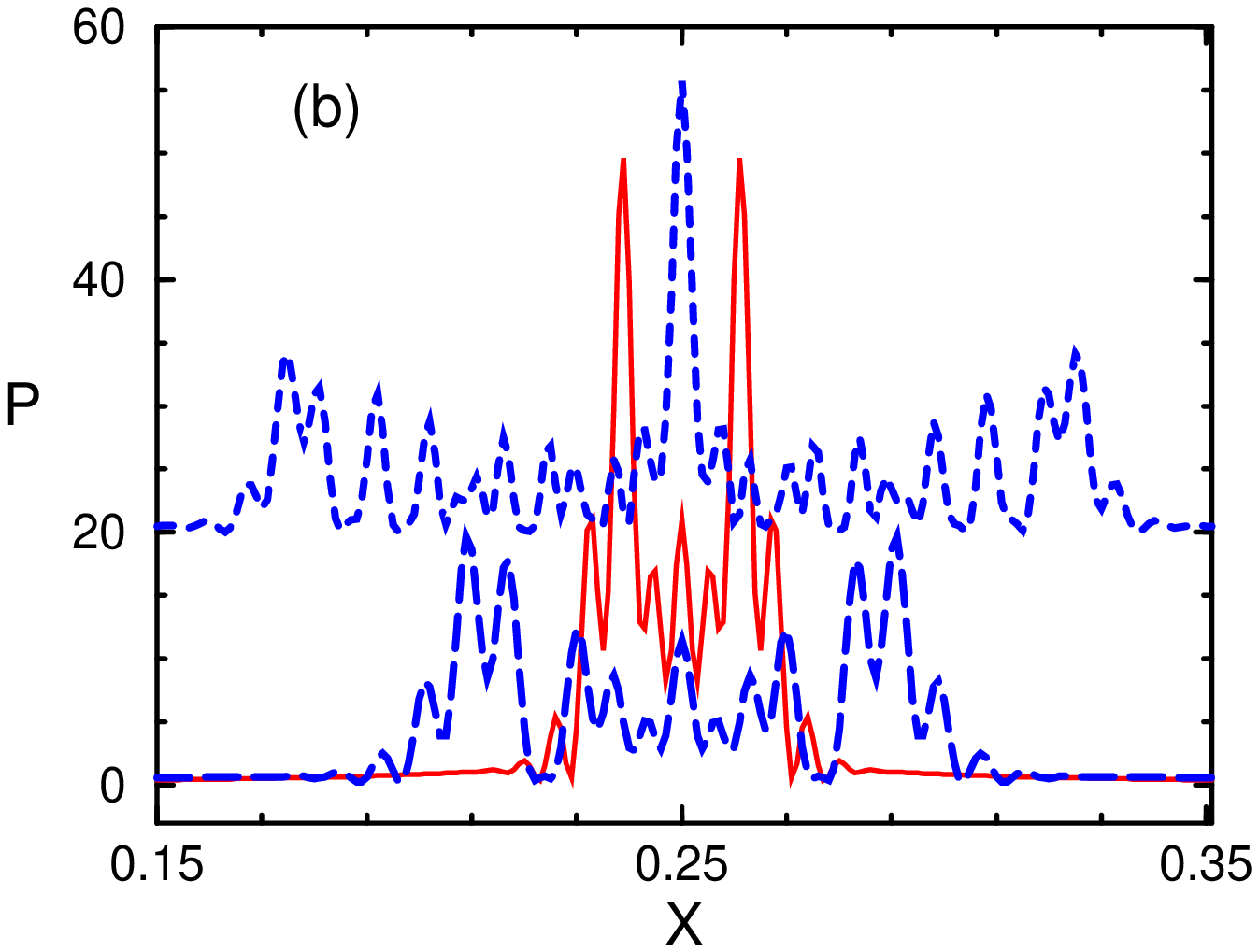}}
\vskip 0.03in
\caption{(Color online) (a) Localization factor of the atomic distribution as
a function of the dimensionless time $t$ for the parameters $\sigma_t = 0.01$,
$\Omega_0 = 4 \times 10^4$, and $\Delta/\Omega_0 = -0.125$ (solid curve),
$\Delta/\Omega_0 = 0.125$ (dashed curve), $\Delta/\Omega_0 = 1$ (long-dashed curve).
The global minimum values of $L(t)$ are 0.1 (solid curve), 0.37 (dashed curve),
and 0.31 (long-dashed curve). (b) Probability density $(P = |\Psi(x,t)|^2)$
of the atomic distribution at the time $t = t_m$ of the best atomic localization.
The parameters are same as those of (a) with $t_m = 0$ (solid curve),
$t_m = 1.5 \times 10^{-3}$ (dashed curve), and $t_m = 5.3 \times 10^{-3}$
(long-dashed curve). For the sake of comparison, we have displaced the solid curve
by 0.25 units along the $X$ axis and the dashed curve by 20 units along the
$P$ axis. The times $t_m$ correspond to the global minima of the localization
factor in (a).}
\end{figure}

\vspace*{0.05in}
\noindent 
similar in the adiabatic regime of atom focusing and there are no
distinct quantum features in the localization of atoms. The wave
effects in propagation have a limited manifestation in this case due
to the relatively short focal length.

When the light frequency is tuned very close to the atomic resonance
$(|\Delta| \ll \Omega_0)$, the adiabatic potentials, Eq. (\ref{pot})
experience sharp spatial variations along the $x$ coordinate in the
regions of avoided crossings near the nodes of the SW. As a result,
non-adiabatic effects degrade greatly the focusing of atoms. In this
case, the atoms do not follow single adiabatic potential (either
plus or minus in Eq. (\ref{pot})) during interaction with the SW but
rather make random transitions between them in the  regions of
quasi-crossing. As a result, they tend to focus near both the minima
and maxima of the light intensity with no well defined localization
region as shown in Fig. 9. The localization factor defined in Eq.
(\ref{qlocal}) is not 

\begin{figure}[t]
\centerline{ \epsfxsize=185 pt \epsfbox{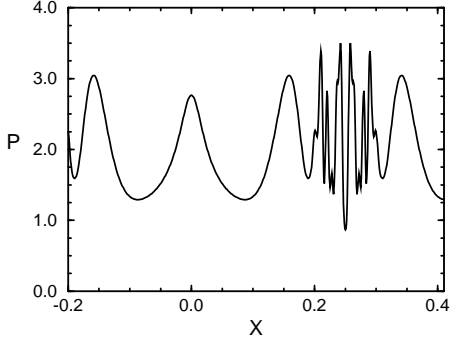}}
\vskip 0.03in
\caption{Probability density $(P = {|\Psi(x,t)|}^2)$ of the atomic 
distribution for the parameters $\sigma_t = 0.01$, 
$\Omega_0 = 2 \times 10^4$, $t=0$, and $\Delta/\Omega_0 = 10^{-2}$.}
\end{figure}
\vspace*{0.05in}
\noindent
appropriate to characterize the atomic distribution for this case.

\section{monte carlo simulations}
We have so far simplified the classical and quantum treatments of
the problem by assuming that the atoms do not change their internal
states by spontaneous emissions during interaction with the light.
This is strictly valid only if $\Gamma \sigma_t \ll 1$, where
$\Gamma$ is the decay rate of the excited atomic level. However, for
chromium atoms which we consider, the quantity $\Gamma \sigma_t$
becomes of the order of 0.1 and 2 in the thin $(\sigma_t = 0.0006)$
and thick $(\sigma_t = 0.01)$ lens regimes, respectively. Therefore,
it is necessary to study the extent to which spontaneous emission
degrades the focusing performance of the adiabatic light potentials.

The effects of spontaneous emission on the focusing of atoms are
two-fold. First, it may modify the trajectories of atoms by
imparting momentum kicks in random directions. Secondly, the dressed
internal state of the atom changes due to spontaneous emissions
resulting in random jumps between the plus and minus adiabatic
potentials, Eq. (\ref{pot}) which the atom experiences. To model all
these features, we employ the quantum Monte Carlo wave function
(MCWF) simulations \cite{{monte1},{monte2}} for the description of
atoms emitting spontaneous photons and moving in a SW light field.
In this approach, the atomic density $P$ is calculated by forming
$N$ realizations of quantum trajectories $\psi_{e,g}^{s}(x,t)$ and
then averaging over them :
\begin{equation}
P \equiv {|\Psi(x,t)|}^2 = \frac{1}{N} \sum_{s=1}^{N} [{|\psi_e^s(x,t)|}^2 + {|\psi_g^s(x,t)|}^2]
~. \label{ensemble}
\end{equation}
Each quantum trajectory $\psi_{e,g}^{s}(x,t)$ consists of a set of
deterministic Hamiltonian  evolution periods interrupted by quantum
collapses. The Hamiltonian evolution is governed by a non-Hermitian
Hamiltonian $H_{eff} = H(t) - i \hbar (\Gamma/2) |e\rangle \langle
e|$, where the term $H(t)$ describes the coherent dynamics of the
atom-light interaction as given in Eq. (\ref{ham}) and the imaginary
term (containing $\Gamma$) accounts for the spontaneous decay of the
excited atomic level. The collapse of the atomic wave function is
given by the action of the operator
\begin{equation}
C_{k'} = {[\Gamma N(k')]}^{1/2} \exp(-i k' x)~|g\rangle \langle e|~~, \label{collapse} \\
\end{equation}
where
\begin{equation}
N(k') = (3/8k) \left[ 1 + {\left(\frac{k'}{k}\right)}^{2} \right]~, \label{dist}
\end{equation}
is the normalized probability density for the distribution of the
spontaneously emitted photons with momentum component $\hbar k'$
along the SW  direction ($x$ axis).

When there is no spontaneous emission event, the internal states of
the atom are coupled only by stimulated processes. The atomic wave
function at time $t$ is given by a Fourier series [cf. Eq.
(\ref{expand})]
\begin{eqnarray}
\psi_e(x,t) &=& \sum_{n=-\infty}^{n=\infty} C^{e}_n(t,t_i)~e^{i[p_0 + (2n+1)\hbar k]x/\hbar}~,
\nonumber \\ 
\psi_g(x,t) &=& \sum_{n=-\infty}^{n=\infty} C^{g}_n(t,t_i)~e^{i[p_0 + 2n\hbar k]x/\hbar} ~.
\label{wavefn}
\end{eqnarray}
where $p_0$ is the momentum of the atom along the SW direction at the initial time $t_i$.
The Fourier coefficients defined above depend on the initial time $t_i$ and evolve with time
$t$ (until a spontaneous emission takes place) as governed by the non-Hermitian Hamiltonian
$H_{eff}$ :
\begin{eqnarray}
i \frac{d}{dt} C^{e}_n(t) &=& \left\{\frac{{[p_o + (2n+1)\hbar k]}^{2}}{2 m\hbar}  -
\frac{\Delta}{2} - \frac{i \Gamma}{2}\right\} C^{e}_n(t) \nonumber \\
&& + \frac{\Omega(t)}{4}(C^{g}_{n}(t) + C^{g}_{n+1}(t))~, \nonumber \\
i \frac{d}{dt} C^{g}_n(t) &=& \left\{\frac{{[p_o + 2n \hbar k]}^{2}}{2 m\hbar} + \frac{\Delta}{2}
\right\} C^{g}_n(t) \label{coeff} \\
&& + \frac{\Omega(t)}{4}(C^{e}_{n}(t) + C^{e}_{n-1}(t))~. \nonumber  
\end{eqnarray}
If a spontaneous emission from the atom takes place at the time $t$,
the momentum $\hbar k'$ of the spontaneously emitted photon is
chosen randomly according to the probability law $N(k')$ [Eq.
(\ref{dist})] and the collapse of the atomic wave function to the
ground state is carried out with the operator Eq. (\ref{collapse})
as follows
\begin{eqnarray}
C^{g}_n(t) &=& C^{e}_n(t)\Bigg{/}\sqrt{\left(\lambda/2\right)\sum_n {|C^{e}_n(t)|}^2}~,
\nonumber \\
C^{e}_n(t) &=& 0~, \label{recoil} \\
p_0 &\rightarrow& p_0 + \hbar k - \hbar k'~. \nonumber
\end{eqnarray}

\begin{figure}[t]
\centerline{ \epsfxsize=185 pt \epsfbox{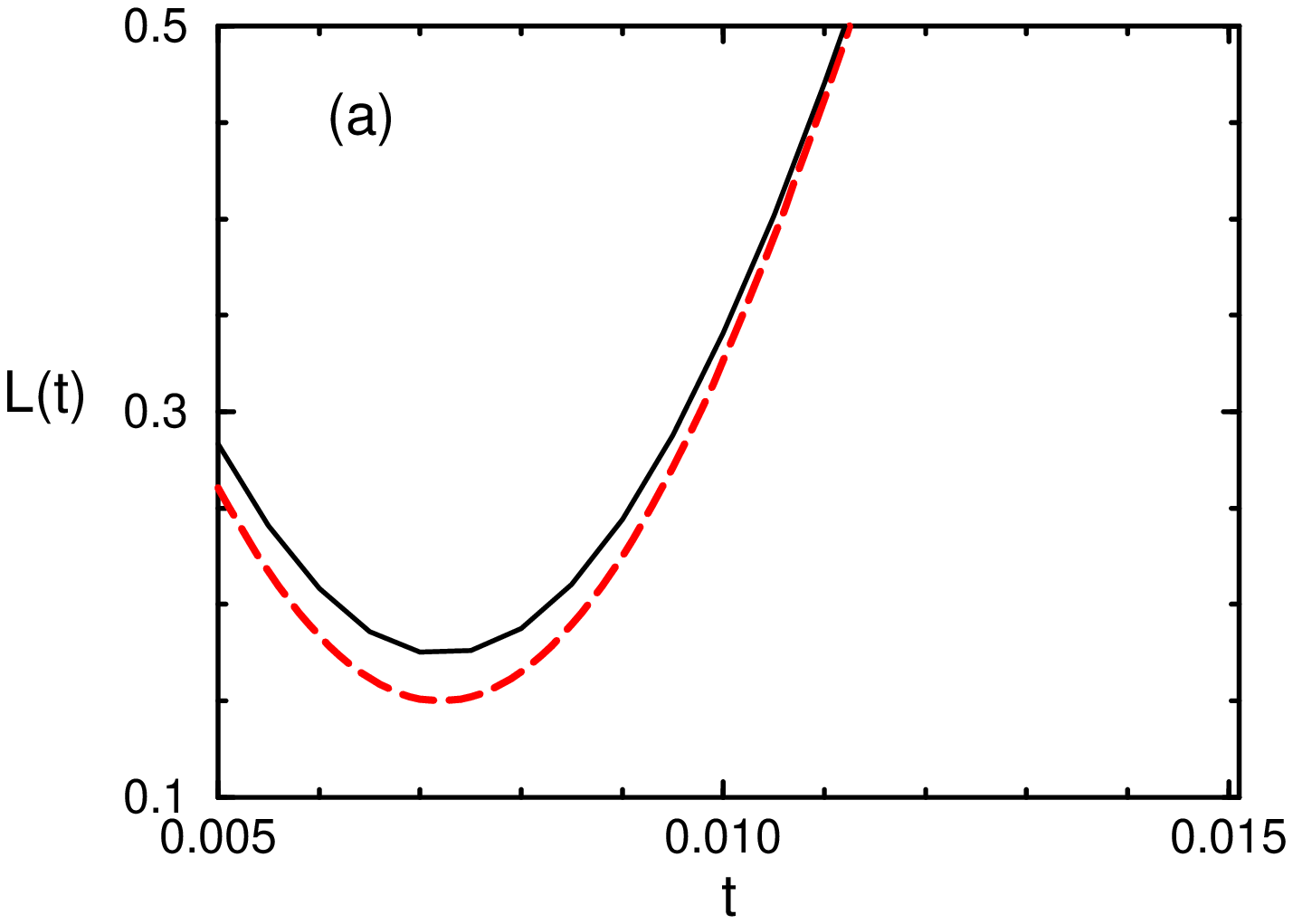}}
\vskip 0.08in \centerline{\epsfxsize=185 pt \epsfbox{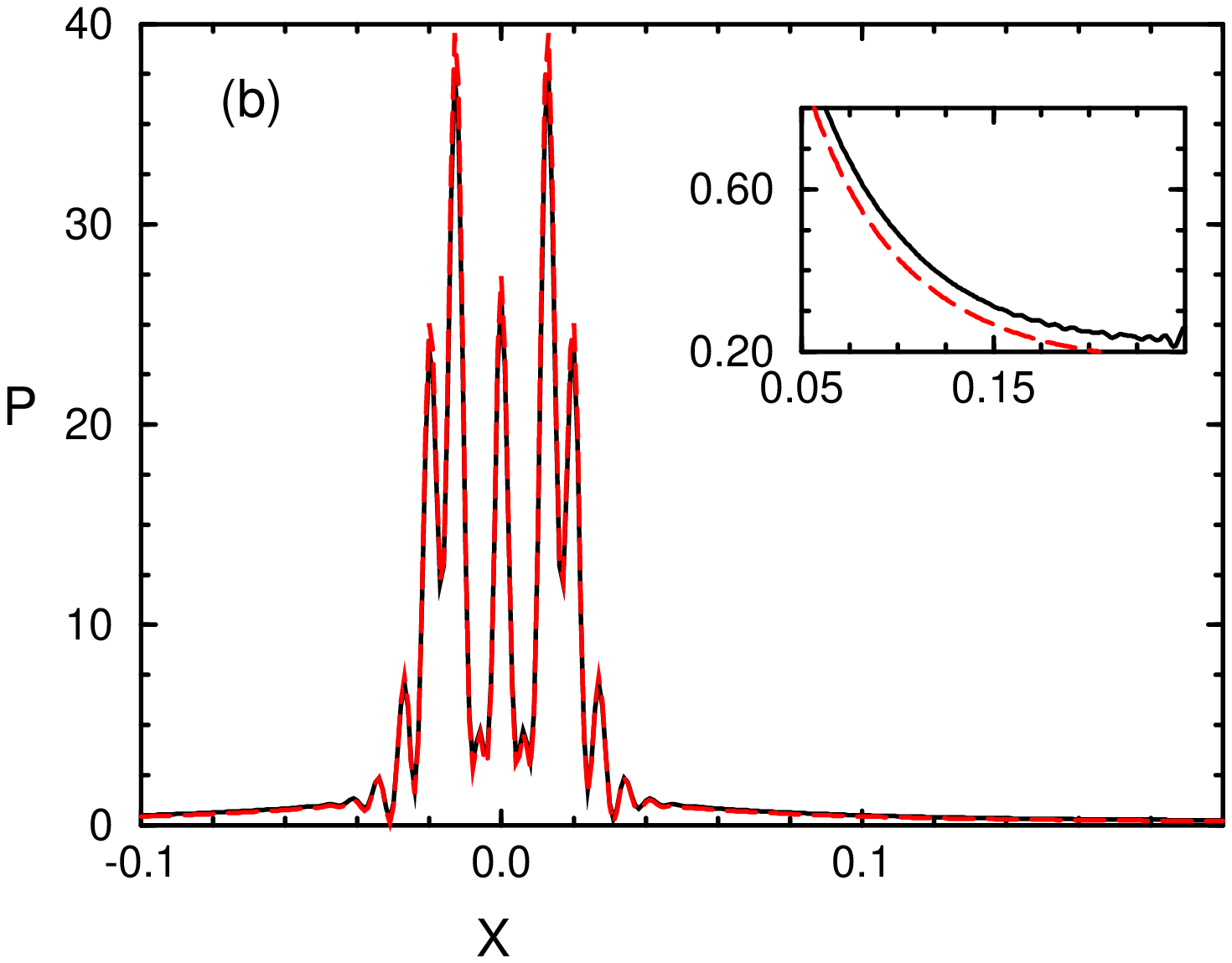}}
\vskip 0.03in
\caption{(Color online) (a) Localization factor of the atomic distribution
as a function of the dimensionless time $t$ around its global minimum for the parameters
$\sigma_t = 0.0006$, $\Omega_0 = 1.92 \times 10^5$, $\Delta/\Omega_0 = -0.125$, and
$\Gamma = 238$ (solid curve), $\Gamma = 0$ (dashed curve). The minimum values of
$L(t)$ are 0.175 (solid curve) and 0.15 (dashed curve). (b) Probability density
$(P = {|\Psi(x,t)|}^2)$ of the atomic distribution at the time $t = t_m$ of
the best atomic localization. The parameters are same as those of (a) with
$t_m = 7 \times 10^{-3}$. The solid curve is almost indistinguishable from the
dashed curve on the scale shown. For clarity, the close-up to the right of
origin $(x=0)$ is shown in the inset. }
\end{figure}

\vspace*{0.18in}
In the MCWF simulations \cite{monte2}, the random moment $t$ at
which the spontaneous emission takes place is chosen when the
decaying total norm of the atomic wave function reaches the value of
$1-\varepsilon$, where $\varepsilon \in [0,...1]$ is a random number
uniformly distributed between 0 and 1. The moment of emission is
determined by solving the equation
\begin{equation}
1 - (\lambda/2) \sum_{n=-\infty}^{n=\infty}
\left[{|C^{e}_n(t,t_i)|}^2 + {|C^{g}_{n}(t,t_i)|}^2 \right] =
\varepsilon~. \label{condn}
\end{equation}
We take the initial time $t_i$ to be $-5 \sigma_t$ (instead of
$-\infty$) at which the Rabi frequency $\Omega(x,t)$ [Eq.
(\ref{rabi})] drops significantly compared to its peak value. We
assume, as before, the initial condition in the form of an atomic
plane wave in the ground state with zero  transverse momentum $(p_0
= 0)$. The norm  ${|\Psi(t)|}^2 = (\lambda/2) \sum_n
\left[{|C^{e}_n(t,t_i)|}^2 + {|C^{g}_{n}(t,t_i)|}^2 \right]$ of the
wave function
\begin{figure}[t]
\centerline{ \epsfxsize=185 pt \epsfbox{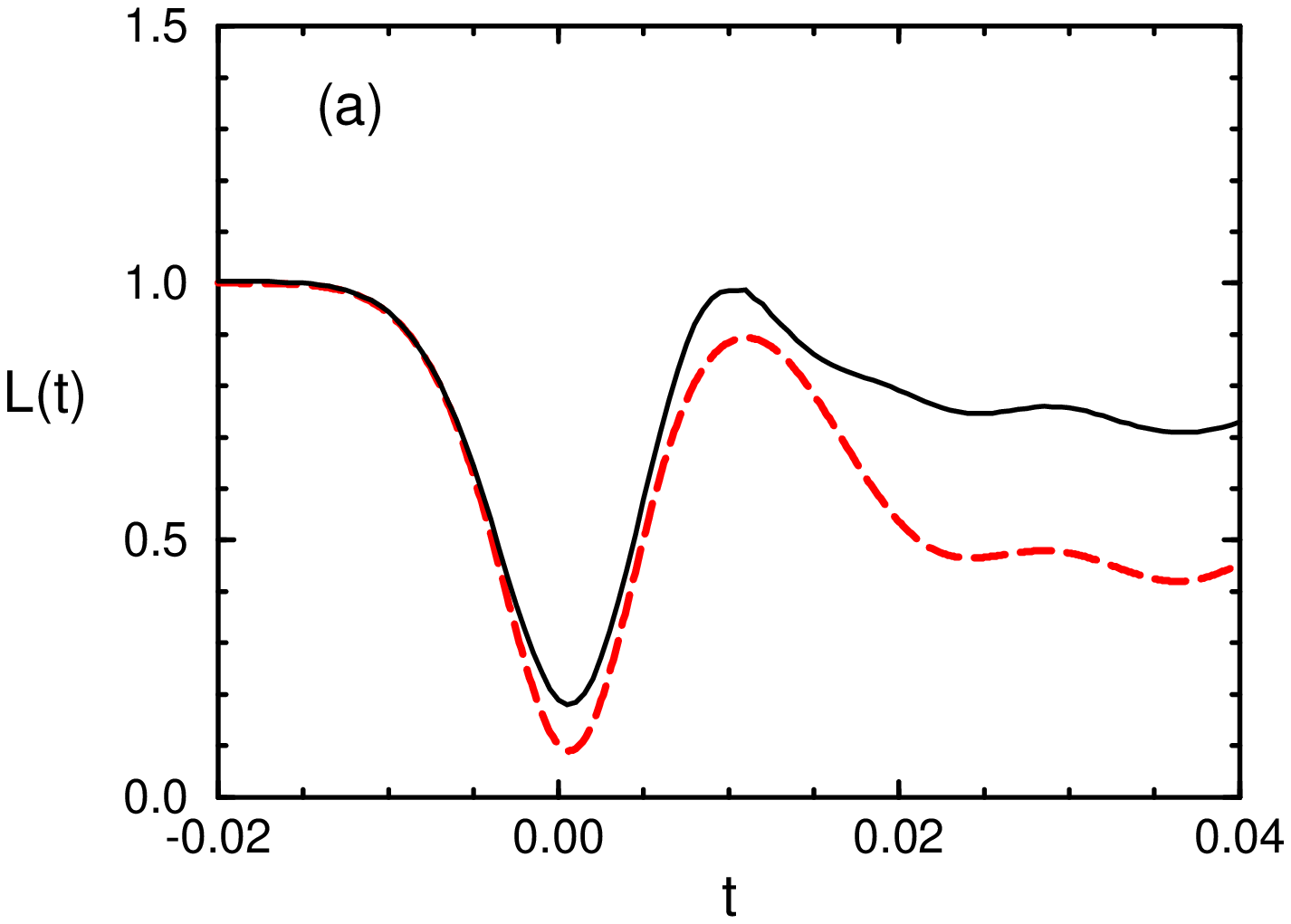}}
\vskip 0.08in \centerline{\epsfxsize=185 pt \epsfbox{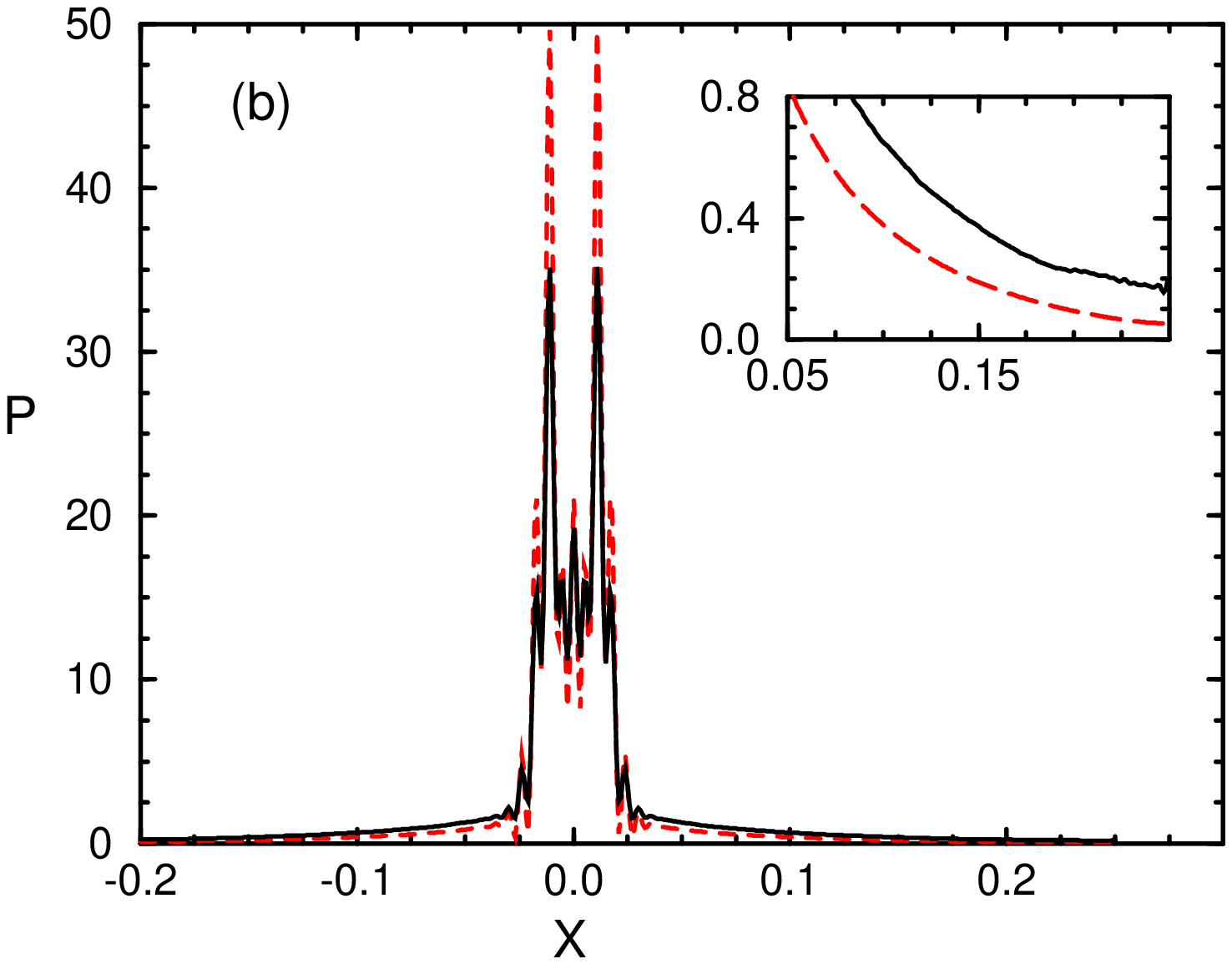}}
\vskip 0.03in
\caption{(Color online) (a) Localization factor of the atomic distribution as a
function of the dimensionless time $t$ around its global minimum for the parameters
$\sigma_t = 0.01$, $\Omega_0 = 4 \times 10^4$, $\Delta/\Omega_0 = -0.125$, and
$\Gamma = 238$ (solid curve), $\Gamma = 0$ (dashed curve). The minimum values of
$L(t)$ are 0.18 (solid curve) and 0.1 (dashed curve). (b) Probability density
$(P = {|\Psi(x,t)|}^2)$ of the atomic distribution at the time $t = t_m$ of
the best atomic localization. The parameters are same as those of (a) with
$t_m = 0$. For clarity, the close-up to the right of origin $(x=0)$ is shown in the
inset.}
\end{figure} 

\vspace*{0.1in}
\noindent
is then obtained by solving Eqs. (\ref{coeff}) until
the time $t = t_{MC}$ at which Eq. (\ref{condn}) is fulfilled. At
this moment, we collapse the atomic wave function to the ground
state and add a recoil momentum to the atom [see Eqs.
(\ref{collapse}) and (\ref{recoil})]. After the collapse, the random
number $\varepsilon$ is renewed and the process gets repeated by
solving again Eq. (\ref{condn}) starting from the new initial time
$t_i = t_{MC}$ and the new values for the initial Fourier
coefficients. This procedure gets continued until the final time $t$
at which the calculation of the deposited atomic density should be
performed. This results in single quantum trajectory
$\psi_{e,g}^{s}(x,t)$ obtained by normalizing the wave function Eqs.
(\ref{wavefn}) at the final time $t$. An ensemble average of many
such trajectories [Eq. (\ref{ensemble})] is statistically equivalent
to the solution of the density matrix equations.

In the limit, when the detuning of the light frequency is relatively
large $(|\Delta|\gg\Omega_0)$, the atoms primarily evolve in their
ground state during interaction with the light. In this case, the
degradation of the atom focusing due to spontaneous emission  is
negligible as reported in earlier studies \cite{{berman},{arun}}.
However, if a near-resonant light is used to focus the atomic beam,
the spontaneous emissions may broaden significantly the atomic
distribution because of the random atomic recoils and fluctuations
of the focusing potentials discussed above. These effects are
especially important in the atom focusing by red-detuned light
fields, since, in this case, the focusing occurs near the light
intensity maxima resulting in substantial population of excited
atoms. In order to measure this quantitatively, we again use the
localization factor Eq. (\ref{qlocal}) with the atomic density
${|\Psi(x,t)|}^2$ given by Eq. (\ref{ensemble}). We find that an
average of $N = 5000$ quantum trajectories is sufficient to give a
statistical error below $2\%$ of the mean value in Eq.
(\ref{ensemble}). The results are shown with $(\Gamma = 238)$ and
without $(\Gamma = 0)$ spontaneous decay of the atoms for the
thin-lens (Fig. 10) and thick-lens (Fig. 11) focusing regimes. It is
seen from the graphs that spontaneous emission does not shift
 the optimal time $t = t_m$ for the global minimum of
the localization factor. But, it reduces the atomic density and
increases the background significantly in the thick-lens focusing of
atoms.

\vspace*{-0.1in}
\section{summary}
\vspace*{-0.17in}
In this paper, we studied the spatial focusing of an atomic beam by
a near-resonant SW light in the context of atom lithography. The
problem was treated both classically and quantum-mechanically using
the optimal squeezing approach  suggested recently \cite{arun} for
the efficient focusing of atomic beams.  Both thin- and thick-lens
regimes were considered  with blue- as well as red-detuned light
fields. In the thick-lens regime, we have shown that a red-detuned
light focuses the atoms better in comparison with a blue-detuned
one and the optimal quantum localization of atoms follows the
classical scenario. In the thin-lens regime, we have shown that
quantum-mechanical effects give rise to new localization mechanisms
which are more effective than the classical ones. Finally, we
considered the role of non-adiabatic transitions between the dressed
atomic states and the effects of spontaneous emission of atoms on 
the quality of the deposition profile. 

\vspace*{-0.15in}
\section{acknowledgements}
\vspace*{-0.17in}
We acknowledge fruitful discussions with D. Meschede, K. M$\o$lmer, and
T. Pfau. This work was supported by the German-Israeli Foundation (GIF) for
Scientific Research and Development.

\end{multicols}

\begin{references}
\bibitem{review1} M. K. Oberthaler and T. Pfau, J. Phys. : Condens. Matter, {\bf 15}, R233 (2003).
\bibitem{review2} D. Meschede and H. Metcalf, J.Phys.D: Appl. Phys. {\bf 36}, R17 (2003).
\bibitem{timp} G. Timp, R.E. Behringer, D.M. Tennant, J.E. Cunningham, M. Prentiss,
and K.K. Berggren, Phys. Rev. Lett. {\bf 69}, 1636 (1992).
\bibitem{prentis} K.K. Berggren, M. Prentiss, G.L. Timp, and R.E. Behringer,
J. Opt. Soc. Am. B {\bf 11}, 1166 (1994).
\bibitem{mcc1} J. J. McClelland, J. Opt. Soc. Am. B {\bf 12}, 1761 (1995).
\bibitem{berman} J. L. Cohen, B. Dubetsky, and P. R. Berman, Phys. Rev. A {\bf 60}, 4886 (1999).
\bibitem{olsen} M. K. Olsen, T. Wong, S. M. Tan, and D.F. Walls, Phys. Rev. A {\bf 53}, 3358 (1996).
\bibitem{lee} C. J. Lee, Phys. Rev. A {\bf 61}, 063604 (2000).
\bibitem{sodium} V. Natarajan, R. E. Behringer, and G. Timp, Phys. Rev. A {\bf 53}, 4381 (1996);
R.E. Behringer, V. Natarajan, and G. Timp, Opt. Lett. {\bf 22}, 114 (1997).
\bibitem{mcc2} J.J. McClelland, R.E. Scholten, E.C. Palm, and R.J. Celotta, Science {\bf 262},
877 (1993); W.R. Anderson, C.C. Bradley, J.J. McClelland, and R.J. Celotta, Phys. Rev. A
{\bf 59}, 2476 (1999).
\bibitem{chr} U. Drodofsky, J. Stuhler, B. Brezger, T. Schulze, M. Drewsen, T. Pfau,
and J. Mlynek, Microelectron. Eng. {\bf 35}, 285 (1997); Th. Schulze, B. Brezger, P.O. Schmidt,
R. Mertens, A. S. Bell, T. Pfau, and J. Mlynek, {\it ibid.} {\bf 46}, 105 (1999);
Th. Schulze, T. M\"{u}ther, D. J\"{u}rgens, B. Brezger, M.K. Oberthaler, T. Pfau, and
J. Mlynek, Appl. Phys. Lett. {\bf 78}, 1781 (2001).
\bibitem{alu} R. W. McGowan, D. M. Giltner, and S. A. Lee, Opt. Lett. {\bf 20}, 2535 (1995).
\bibitem{ytt} R. Ohmukai, S. Urabe, and M. Watanabe, Appl. Phys. B: Lasers Opt. {\bf 77}, 415 (2003).
\bibitem{iro} E. te Sligte, B. Smeets, K. M. R. van der Stam, R. W. Herfst,
P. van der Straten, H. C. W. Beijerinck, and K. A. H. van Leeuwen, Appl. Phys. Lett.
{\bf 85}, 4493 (2004); G. Myszkiewicz, J. Hohlfeld, A. J. Toonen, A. F. Van Etteger,
O. I. Shklyarevskii, W. L. Meerts, Th. Rasing, and E. Jurdik, {\it ibid.} {\bf 85},
3842 (2004).
\bibitem{twod} R. Gupta, J.J. McClelland, Z.J. Jabour, and R.J. Celotta, Appl. Phys. Lett.
{\bf 67}, 1378 (1995); U. Drodofsky, J. Stuhler, T. Schulze, M. Drewsen, B. Brezger, T. Pfau,
and J. Mlynek, Appl. Phys. B: Lasers Opt. {\bf 65}, 755 (1997).
\bibitem{arun} R. Arun, I. Sh. Averbukh, and T. Pfau, Phys. Rev. A {\bf 72}, 023417 (2005);
M. Leibscher and I. Sh. Averbukh, {\it ibid.} {\bf 65}, 053816 (2002).
\bibitem{resonant} D. J\"{u}rgens, A. Greiner, R. St\"{u}tzle, A. Habenicht, Edwin te Sligte,
and M.K. Oberthaler, Phys. Rev. Lett. {\bf 93}, 237402 (2004).
\bibitem{dress} C. Cohen-Tannoudji, J. Dupont-Roc, G. Grynberg, {\it Atom-Photon Interaction}
(John Wiley, New York, 1992).
\bibitem{adiab} More precisely, the condition for adiabatic following is
$|\langle E_j(t)|\dot{H}(t)|E_i(t)\rangle| \ll {|E_j(t) - E_i(t)|}^{2}$, where $|E_{i,j}(t)\rangle$
are the time-dependent dressed states (with eigenenergies $E_{i,j}(t)$) of the Hamiltonian $H(t)$.
However, an approximate and simplified expression can be obtained by evaluating it using the bare
atomic states and energies.
\bibitem{pfau} The first experimental demonstration of atom focusing using a thin SW lens was
reported by T. Sleator, T. Pfau, V. Balykin, and J. Mlynek, Appl. Phys. B: Photophys. Laser Chem.
{\bf 54}, 375 (1992).
\bibitem{norm} In all figures, the probability density of the atomic distribution is normalized
such that its integral over a standing wave period $(\lambda/2)$ is unity.
\bibitem{jurgens} D. J\"{u}rgens, Ph. D. dissertation, Universit\"{a}t Konstanz, Germany, 2004.
\bibitem{ground} Note that the atom evolves primarily in its ground internal state outside the
light region if the atom adiabatically follows the potential (\ref{pot}).
\bibitem{monte1} K. M$\o$lmer, Y. Castin, and J. Dalibard, J. Opt. Soc. Am. B {\bf 10}, 524 (1993);
R. Blatt, W. Ertmer, P. Zoller, and J.L. Hal, Phys. Rev. A {\bf 34}, 3022 (1986); R. Dum, P. Zoller,
and H. Ritsch, {\it ibid.} {\bf 45}, 4879 (1992).
\bibitem{monte2} M.D. Hoogerland {\it et al.,} Phys. Rev. A {\bf 54}, 3206 (1996).
\end{references}
\end{document}